\documentclass[aps,prb,reprint,nofootinbib,twocolumn,superscriptaddress,showpacs,showkeys,longbibliography]{revtex4-1}
\usepackage{eurosym}
\usepackage{amsmath,amssymb,amstext}
\usepackage[usenames,dvipsnames]{color}
\usepackage{graphicx}
\usepackage{braket}
\usepackage{natbib}
\usepackage{comment}
\usepackage{dcolumn}
\usepackage[english]{babel}
\usepackage{wasysym}
\usepackage[colorlinks,bookmarks=false,citecolor=blue,linkcolor=red,urlcolor=blue]{hyperref}

\usepackage{graphicx}
\usepackage{dcolumn}
\usepackage{bm}

\begin{document}

\title{Self-Decelerating Bright Exciton-Polariton Solitons in Bound-State-in-Continuum Microcavities}
\author{Xingran Xu}
\email{thoexxr@hotmail.com}
\affiliation{School of Optoelectronic Information and Physical Science, Jiangnan University, Wuxi 214122, China}

\author{Chunyu Jia}
\affiliation{College of Physical Science and Technology, Bohai University, Jinzhou 121013, China}

\date{\today}

\begin{abstract}
We theoretically investigate the formation and dynamics of bright exciton-polariton solitons within systems engineered to support Bound States in the Continuum. By employing a driven-dissipative Gross-Pitaevskii equation coupled with a rate equation for the excitonic reservoir, we demonstrate that BICs provide a robust platform for stabilizing the condensate against radiative decay. Utilizing a Lagrangian variational approach, we derive analytical expressions describing the trajectory and velocity of these bright solitonic excitations. Notably, we find that the propagation of these BIC-engineered solitons exhibits a distinct self-deceleration, eventually bringing them to a halt at a final position dictated by the initial conditions and intrinsic system parameters. Furthermore, we analyze the dynamical stability of these solitons. Our findings offer valuable insights into the manipulation of polaritonic flows in non-Hermitian systems.
\end{abstract}

\maketitle


\section{Introduction\label{sec:level1} }

Exciton-polaritons, hybrid light-matter quasiparticles arising from the strong coupling between excitons and cavity photons, have garnered significant attention over the past two decades due to their unique properties and promising applications in all-optical signal processing and quantum information technologies \cite{Kavokin2007,Deng2010,Carusotto2013}. These bosonic quasiparticles possess exceptionally low effective masses, typically on the order of $10^{-5}$ times the free electron mass, which facilitates the observation of macroscopic quantum phenomena at elevated temperatures \cite{Kasprzak2006,Byrnes2014}. The intrinsic gain and loss mechanisms inherent to polaritons make them an ideal platform for exploring non-Hermitian physics, including exceptional points \cite{EPGAO,EPJan,EPChen_2017}, non-reciprocal transport \cite{Yang_2023,Liang_2025,del_Valle_Inclan_Redondo_2024,bjb9-wtrb}, and the non-Hermitian skin effect \cite{Xu2021,XU2022,Bao2023,Xu_2025}. Furthermore, the nonlinear interactions stemming from the excitonic component give rise to a rich variety of collective behaviors, such as Bose-Einstein condensation, superfluidity, and the formation of localized solitary waves \cite{Amo2009,Amo2011,Nardin2011}.

Among the various nonlinear excitations in polariton systems, bright exciton-polariton solitons stand out as particularly intriguing subjects of study. These self-localized wave packets maintain their shape during propagation by balancing the effects of dispersion and nonlinearity \cite{Egorov2009,Sich2012}. However, in conventional semiconductor microcavities, polariton solitons face significant challenges due to the inherently dissipative nature of these systems. The finite lifetime of cavity photons, typically on the picosecond timescale, leads to the rapid decay of the condensate and severely limits the propagation distance of solitonic structures \cite{Gippius2007,Wouters2007,Smirnov2014,Hu2025}. This radiative loss fundamentally constrains the practical application of polariton solitons in integrated photonic devices.

To overcome these limitations, various strategies have been proposed to enhance the stability of polariton condensates against radiative decay. One promising approach involves engineering the dispersion landscape of the system to realize Bound States in the Continuum (BICs)—resonant modes that coexist with extended continuum states yet remain perfectly localized due to symmetry protection or destructive interference effects \cite{Hsu2016,Kodigala2017,Zhen2014}. BICs have recently been demonstrated in photonic systems including photonic crystals \cite{Plotnik2011,Lee2012,Gao2016,Koshelev2019,Koshelev_2020}, waveguide arrays \cite{Yan_2024,Longhi2014,Mukherjee2020}, and microcavities . In the context of exciton-polaritons, BICs offer a particularly attractive platform as they can provide arbitrarily high quality factors while maintaining strong light-matter coupling.

The BIC mechanism in polariton systems manifests as an inverted quadratic dispersion relation near the center of the Brillouin zone, where the radiative decay rate vanishes at zero in-plane momentum and increases quadratically with increasing wavevector \cite{Seet_2025,Wu_2024}. This unique dispersion engineering effectively suppresses the radiative losses for localized condensates with narrow momentum distributions, thereby significantly extending the polariton lifetime. Recent experimental breakthroughs have demonstrated that such intrinsic strong light-matter coupling can be achieved in van der Waals metasurfaces, providing a robust platform to explore these phenomena \cite{Weber_2023}. Understanding how BICs modify the formation criteria, propagation dynamics, and stability of these solitons is therefore essential for harnessing these structures in practical devices, with recent studies on asymmetric hetero-bilayer metasurfaces further highlighting their potential for high-Q sensing applications \cite{PARK2024110191}.

The interplay between BIC physics and nonlinear wave dynamics opens new avenues for controlling the propagation of polaritonic excitations. In conventional dissipative polariton systems, solitons typically exhibit self-acceleration or deceleration depending on the pump configuration and reservoir dynamics \cite{Wangstable,PhysRevB.77.115336,Septembre2024}. The introduction of BIC engineering fundamentally alters this behavior, as the momentum-dependent loss landscape couples the soliton's spatial profile to its decay characteristics. Understanding how BICs modify the formation criteria, propagation dynamics, and stability of bright exciton-polariton solitons is therefore essential for harnessing these structures in practical devices.

In this work, we theoretically investigate the formation and transport dynamics of bright exciton-polariton solitons in systems engineered with Bound States in the Continuum. By employing a driven-dissipative Gross-Pitaevskii equation coupled with a rate equation for the excitonic reservoir, we demonstrate that BICs provide a robust platform for stabilizing the condensate against radiative decay. Using a Lagrangian variational approach with a sech-shaped ansatz, we derive analytical expressions for the trajectory and velocity of these bright solitonic excitations. Our analysis reveals that the propagation of BIC-engineered structures exhibits distinct self-deceleration behavior, eventually halting at a final position determined by the initial momentum and the BIC fitting parameter. Furthermore, we analyze the dynamical stability of these solitons against perturbations in the initial amplitude and reservoir parameters. Our findings provide fundamental insights into the manipulation of polaritonic flows in non-Hermitian systems and establish design principles for BIC-based polaritonic devices with controlled transport property. The remainder of the paper is structured as follows: In Section \ref{secmodel}, we present the BIC model of exciton-polaritons and calculate the dispersion and emission properties of the system. In Section \ref{variational}, we employ the Lagrangian variational approach to derive the equation of motion for the bright soliton's central position. In Section \ref{secinstable}, we analyze the stability of the soliton dynamics. A summary is presented in Section \ref{secconclu}.

\section{Model\label{secmodel} }
We consider that the dynamics of BIC exciton-polaritons under nonresonant pumping can be described by the driven-dissipative Gross-Pitaevskii equation for the time-dependent wavefunction $\Psi$:
\begin{eqnarray}
i\hbar\frac{\partial}{\partial t} \Psi&=&\left[(i\Lambda-1)\frac{\hbar^2}{2m^{*}}\frac{\partial^2}{\partial x^2}+g|\Psi|^2+g_R n_R\right]\Psi \nonumber\\
&+&\frac{i\hbar}{2}\left[Rn_R-\gamma_C\right]\Psi, \label{GP0}
\end{eqnarray}
where $m^{*}=-m$ is the negative effective mass \cite{Wurdack_2023,Septembre2024},  $\Lambda$ is the fitting coefficient from the dispersion \cite{}, $g$ and $g_R$ are the interaction strengths between polaritons and the interactions strengths between polaritons and the reservoir. $\Lambda$ denotes a coefficient obtained through experimental fitting, $\gamma_C$ is the constant decay of the condensate and the $R$ is the stimulated scattering rate of reservoir $n_R(x,t)$ into the condensate. The reservoir can be described by a rate equation
\begin{equation}
\frac{\partial}{\partial t}n_R=P-\left[\gamma_R-R|\Psi|^2\right]n_R,
\end{equation}
where $\gamma_R$ is the decay rate of the reservoir and $P$ is the pump strength of the laser. The threshold of the pump rate is $P_{\text{th}}=\gamma_C\gamma_R/R$. When $P>P_{\text{th}}$, the polariton condensates will exist with the density of $n_0=(P-P_{\text{th}})/\gamma_C$ and a steady reservoir density $n_R^0=\gamma_C/R$. These steady states are established under the equilibrium condition $\partial n_R/\partial t=0$ and balance of the particle number in Eq. (\ref{GP0}).

\begin{figure}
    \centering
    \includegraphics[width=1\linewidth]{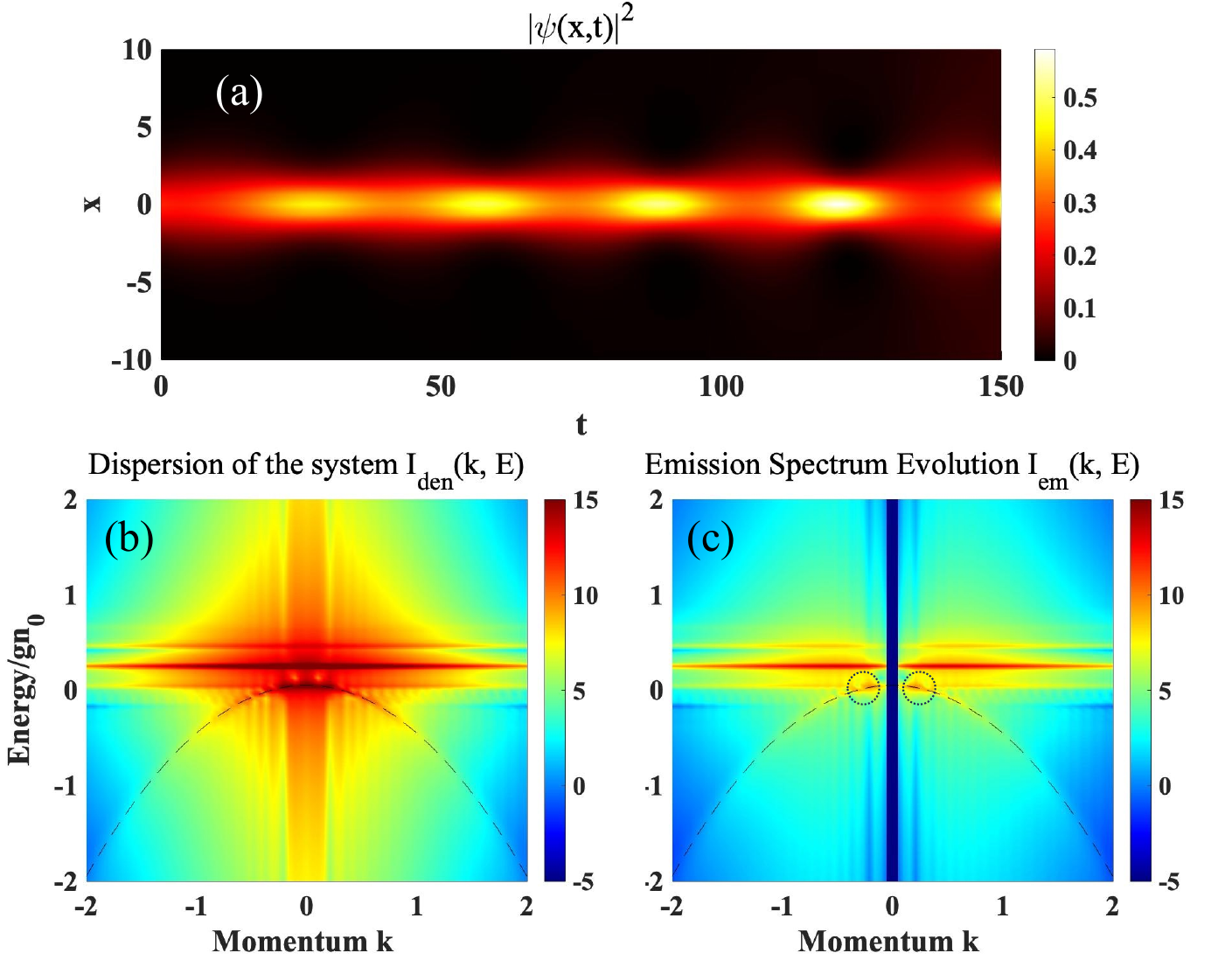}
    \caption{Density distribution $|\psi(x,t)|^2$, along with the dispersion (b) and the emission (c) spectrum of the BIC polaritons. Parameters are: $\eta(0)=0.5$, $k(0)=0$, $\Lambda=-0.2$, $\bar{R}=1.5$, $\bar{\gamma}_C=1$, and  $\bar{\gamma}_R=40$. }
    \label{dispersion}
\end{figure}

To investigate the stability of the condensate, we consider the steady-state regime above the threshold, where the perturbation dynamics are governed by two coupled dimensionless equations \cite{Smirnov2014}:
\begin{eqnarray}
i\frac{\partial}{\partial t} \psi&=&\left[-\frac{i\Lambda-1}{2}\frac{\partial^2}{\partial x^2}+|\psi|^2+\bar{g}_R n_R\right]\psi \nonumber\\
&+&\frac{i}{2} \bar{R} m_R\psi,\\
\frac{\partial}{\partial t}m_R&=&\bar{\gamma}_C\left( 1-|\psi|^2\right)-\bar{\gamma}_Rm_R-\bar{R}|\psi|^2 m_R,
\end{eqnarray}
where, we take $n_R=n_R^0+m_R(x,t)$, $\Psi=\sqrt{n_0}\psi(x,t)$. The spatial coordinate is normalized as $x=x/(\hbar/mc_s)$ with $c_s=\sqrt{gn/m}$ being the local sound velocity of the condensation, and the time unit is $\hbar/gn $. The remaining dimensionless parameters are defined as follows: $\bar{R}=\hbar R/g$, $\bar{g}_R=g_R/g$, $\bar{\gamma}_C=\gamma_C \bar{\gamma}_R/\gamma_R$,and $\bar{\gamma}_R=\bar{R}/(P/P_{\text{th}}-1)$.

Owing to the suppressed radiative decay inherent to the system, BIC solitons exhibit significantly enhanced stability compared to their conventional dissipative counterparts. To characterize these robust localized states, we express the bright soliton solution in the following form:
\begin{equation}
\psi(x,t)=\eta\text{sech}\left[\eta(x-z)\right]\exp\left[ikx +i\phi \right], \label{wf}
\end{equation}
where $\eta=\eta(t)$ is the amplitude of the soliton, $z=z(t)$ is the central position of the solitons, $k=k(t)$ is the velocity, and $\phi=\phi(t)$ is the phase of the soliton.

 \begin{figure*}
    \centering
    \includegraphics[width=1\linewidth]{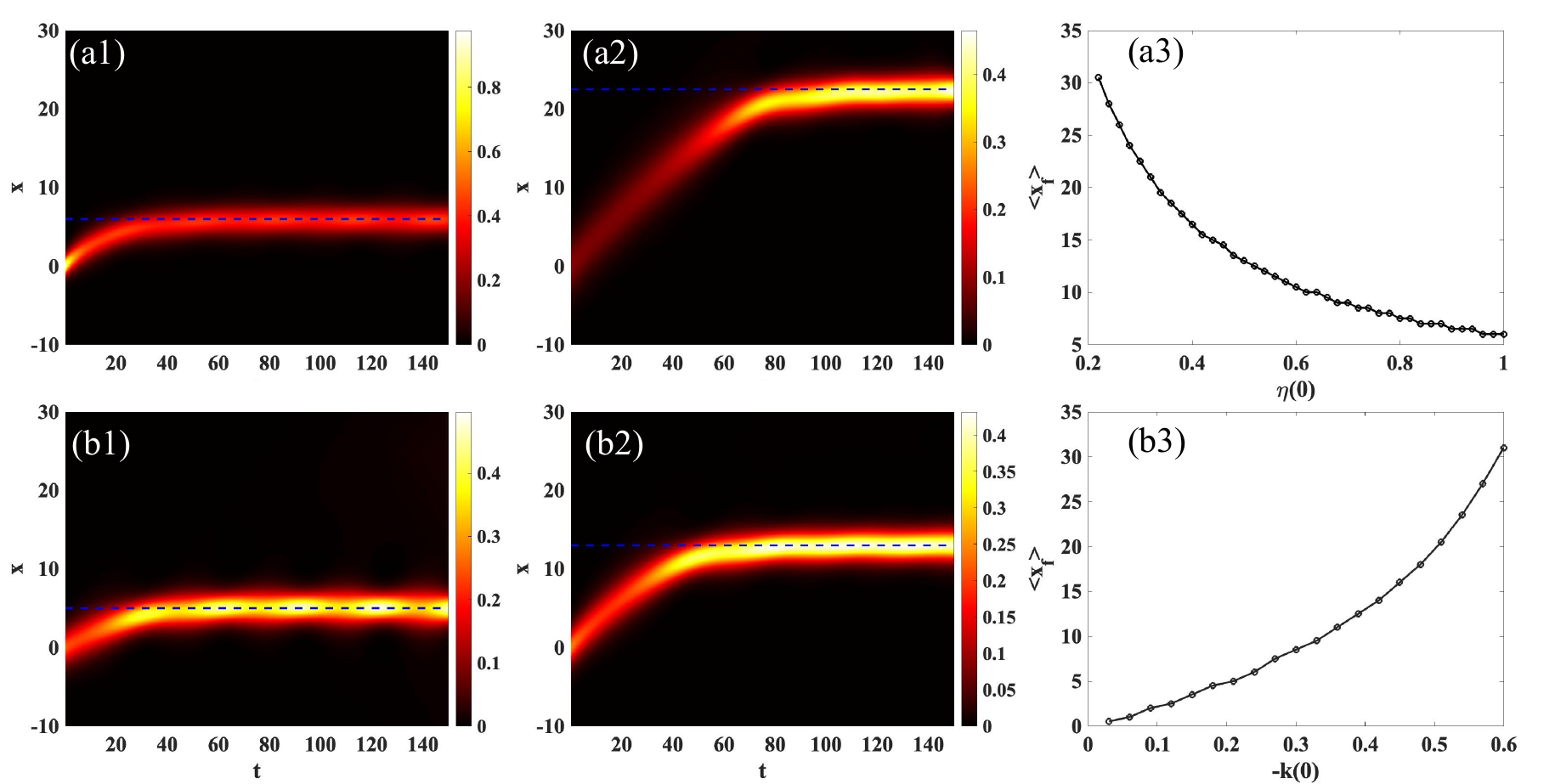}
    \caption{ \textbf{Transport dynamics of BIC polaritons initialized with distinct states.} (a1)-(a2) Density distributions of $|\psi(x,t)|^2$ for initial amplitude $\eta(0)=1$ and $0.3$, respectively, with a fixed initial momentum $k(0)=0.4$.  (a3) Final position $<x_f>$ as a function of the initial amplitude  $\eta(0)$.   (b1)-(b2) Density distributions $|\psi(x,t)|^2$ for initial momenta $k(0)=-0.2$ and $-0.4$, respectively, with a fixed initial amplitude $\eta(0)=0.5$. (b3) Final position as a function of the initial momentum  $k(0)$.  Other parameters are: $\bar{\gamma}_R=40$, $\bar{\gamma}_C=1$, $\bar{R}=1.5$ and $\Lambda=-0.2$. \label{initialchange} }
\end{figure*}

As illustrated in Fig. \ref{dispersion}(a), the bright soliton exhibits a lifetime significantly exceeding that of the condensates. To further analyze the spectral characteristics, the dispersion  $I_{\text{den}}(k,E)$, calculated as the absolute square of the Fourier transform of the condensate density  $\psi(x,t)$, displays an inverted quadratic dependence on the momentum $-k^2/2$ in Fig. \ref{dispersion}(b). The color scale indicates the logarithmic intensity of $I_{\text{den}}(k,E)$, revealing that the majority of the condensate is localized around $k$=0.

The lifetime of polaritons, similar to that of photonic modes, exhibits an inverted quadratic dependence on the wave vector for small $k$. An ideal non-interacting homogeneous condensate would occupy a single state at $k = 0$, thereby benefiting from an exceptionally long lifetime. However, various mechanisms can induce spatial localization of the condensate in real space, which in turn leads to momentum-space broadening and an enhanced overall decay rate (i.e., a reduced lifetime). This contribution can be evaluated as:
\begin{equation}
I_{\text{em}}= \Lambda\frac{\hbar^2k^2}{2m^*}|\psi(k,E)|^2.
\end{equation}
As a result, the BIC branch exhibits zero loss at zero momentum. However, this loss is significantly enhanced at finite momenta, as shown in Fig. \ref{dispersion}(c).  Given that the $Q$ quality factor is determined by the ratio of the real to imaginary parts of the complex frequency, it theoretically diverges (approaches infinity) at $k=0$.

\section{The Lagrange variational approach for the motion of bright solitons \label{variational}}

\begin{figure*}
    \centering
    \includegraphics[width=1\linewidth]{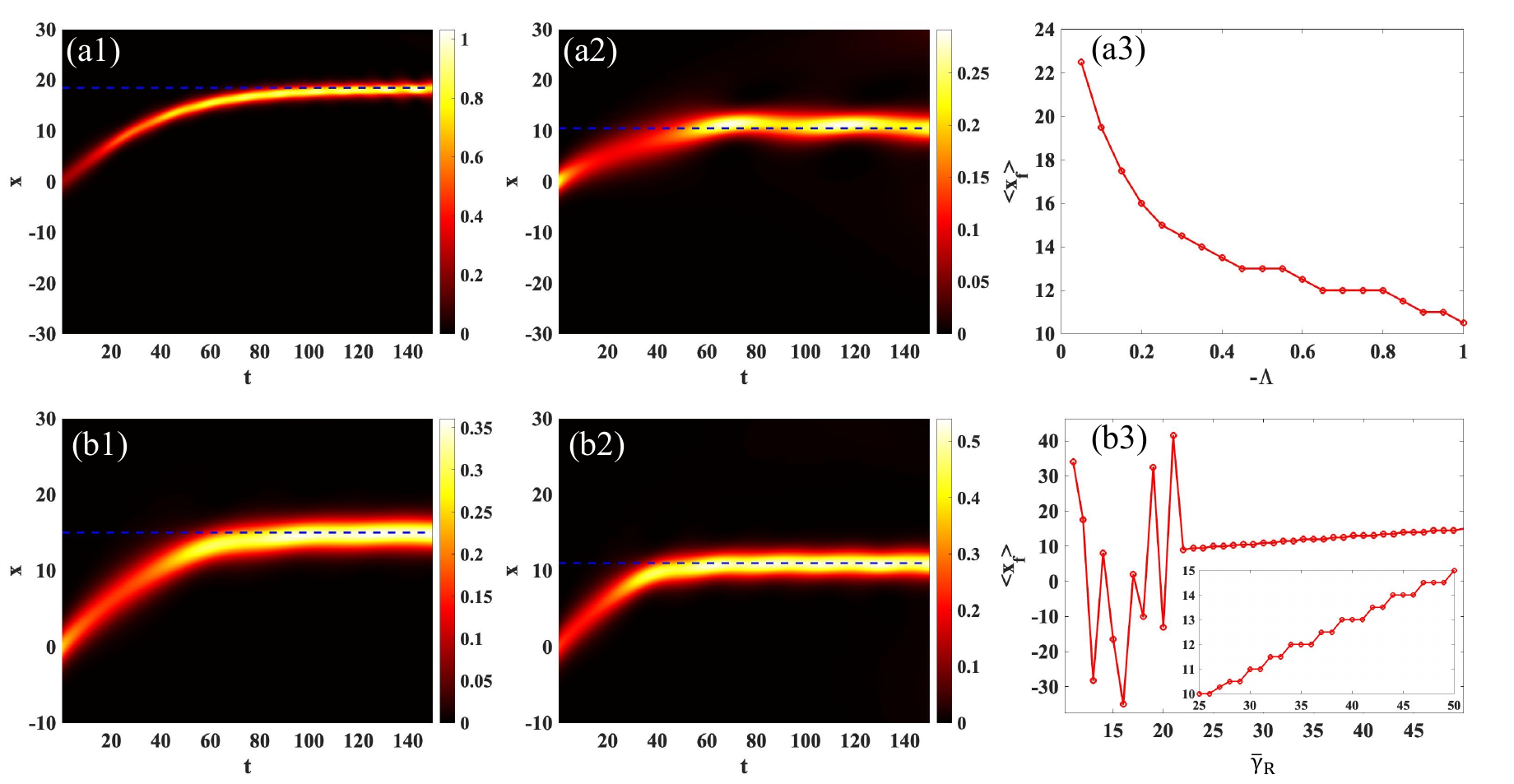}
    \caption{\textbf{Transport dynamics of BIC polaritons with different materials parameters.} (a1)-(a2) Density distributions of $|\psi(x,t)|^2$ for $\Lambda=-0.1$ and $-0.4$, respectively, with a fixed $\bar{\gamma}_R=40$. (a3) Final position $<x_f>$ as a function of $\bar{\Lambda}$.  (b1)-(b2) Density distributions of $|\psi(x,t)|^2$ for $\bar{\gamma}_R=50$ and $30$, respectively, with a fixed $\Lambda=-0.2$. (b3) Final position $<k_f>$ as a function of $\bar{\gamma}_R$. Other parameters are: $\bar{\gamma}_C=1$, $\bar{R}=1.5$, $\eta(0)=0.5$, and $k(0)=-0.4$. \label{conditionchange} }
\end{figure*}

The solitonic nature of these nonlinear excitations—including their characteristic spatial localization—is preserved over extended timescales, primarily governed by the specific system parameters. However, intrinsic dissipation and the BIC mechanism fundamentally dictate the dynamical behavior of the bright solitons. Building on these observations, we demonstrate that within the framework of our model, simple analytical expressions for the velocity and trajectories of these bright solitonic excitations in BIC polariton condensates can be rigorously derived \cite{fsp4-cwh6,Egorov2009}.
Specifically, the interplay between gain, loss, and nonlinearity allows for the stabilization of these structures, enabling precise analytical treatment that confirms the robust propagation of solitons despite the dissipative environment \cite{Ostrovskaya2012,SICH2016908}.

In our following calculation, we just consider the weak pumping condition with $P\approx P_{\text{th}}$ and $\bar{\gamma}_R\gg\bar{\gamma}_C$. Therefore, the equation of the reservoir parts can be approached as the perturbation of the normal nonlinear GP equation with negative mass
\begin{equation}
i\frac{\partial}{\partial t}\psi-\frac{1}{2}\nabla^{2}\psi-\left|\psi\right|^{2}\psi=D\left(\psi\right), \label{eq: Model equations to be solved}
\end{equation}
where, the perturbation is explicitly given by
\begin{equation}
D\left(\psi\right)=-\frac{i\Lambda}{2}\nabla^{2}\psi+\left(\bar{g}_{R}+\frac{i}{2}\bar{R}\right)\frac{\bar{\gamma}_{C}}{\bar{\gamma}_{R}}\psi\left(1-\left|\psi\right|^{2}\right).
\end{equation}

To solve Eq. (\ref{eq: Model equations to be solved}), we employ the Lagrangian variational approach with the variational wavefunction ansatz given in Eq. (\ref{wf}). The time evolution of the variational parameters in Eq. (\ref{wf})  can be obtained via the Euler-Lagrangian equations. Thus, the Euler-Lagrange equations for the variational parameters are given by \cite{Xu_2019,wei2019}: 
\begin{equation}
\frac{\partial L}{\partial q_{i}}-\frac{d}{dt}\left(\frac{\partial L}{\partial\dot{q}_{i}}\right)=\int_{-\infty}^{+\infty}\left[D^{*}\left(\psi\right)\frac{\partial\psi}{\partial q_{i}}+D\left(\psi\right)\frac{\partial\psi^{*}}{\partial q_{i}}\right]dx,  \label{eq: the Euler-Lagrangian equations} 
\end{equation}
with $\dot{q_{i}}\equiv\frac{dq_{i}}{dt}$ and $q_{i}=\eta\left(t\right), z\left(t\right)$, $k\left(t\right)$, $\phi\left(t\right)$.

In Eq. (\ref{eq: the Euler-Lagrangian equations}) , the Lagrangian $L = \int_{-\infty}^{+\infty}\mathcal{L}dx$ is referred to as the average Lagrangian of the equation (\ref{eq: Model equations to be solved}) , where the Lagrangian density $\mathcal{L}$ is given by
\begin{equation}
\mathcal{L}=\frac{i}{2}\left(\psi^{*}\psi_{t}-\psi\psi_{t}^{*}\right)-\mathcal{H}, \label{eq: the Lagrangian density} 
\end{equation}
where the Hamiltonian density is given by $\mathcal{H}=-\frac{1}{2}\left|\psi_{x}\right|^{2}+\frac{1}{2}\left|\psi\right|^{4}$. Substituting the wave function  (\ref{wf}) into the Lagrangian density $\mathcal{L}$ , we obtain the final Lagrangian $L = \int_{-\infty}^{+\infty}\mathcal{L}dx$:
\begin{equation}
L=-2\eta\left(z\dot{k}+\dot{\phi}\right)+\frac{1}{3}\eta\left(\eta^{2}+3k^{2}\right)-\frac{2}{3}\eta^{3}. \label{eq: the Lagrangian} 
\end{equation}

According to the Euler-Lagrange equation for nonequilibrium systems, i.e., Eq. (\ref{eq: the Euler-Lagrangian equations}). By substituting Eq. (\ref{eq: the Lagrangian}) into Eq. (\ref{eq: the Euler-Lagrangian equations}), we obtain the equations of motion for the variational parameters $\eta\left(t\right)$, $z\left(t\right)$, $k(t)$ and $\phi\left(t\right)$ in  Eq. (\ref{wf}) as:
\begin{subequations}
\begin{eqnarray}
\dot{\eta} & = & \frac{\bar{R}\bar{\gamma}_{C}}{\bar{\gamma}_{R}}\eta\left(1-\frac{2}{3}\eta^{2}\right)+\frac{1}{3}\Lambda\eta\left(\eta^{2}+3k^{2}\right), \label{Dy1}\\ 
\dot{z} & = & -k, \label{Dy2}\\ 
\dot{k} & = & \frac{2}{3}\Lambda\eta^{2}k,  \label{Dy3}\\
\dot{\phi} & = & \frac{1}{2}\left(-\eta^{2}+k^{2}\right)-2\Lambda k z\eta^{2}-\frac{\bar{g}_{R}\bar{\gamma}_{C}}{\bar{\gamma}_{R}}\left(1-\eta^{2}\right), \label{Dy4}
\end{eqnarray}
\end{subequations}
which describes the time evolution of the BIC bright solitons in the exciton-polaritons.

An exact analytical solution to these coupled equations is intractable. Nevertheless, we may assume that the soliton amplitude $\eta$ attains a stationary value for $\eta_s$. As shown in Figs  \ref{initialchange}(a1)-(a2) and (b1)-(b2), BIC solitons exhibit self-deceleration, which distinguishes them from the self-acceleration observed under normal conditions \cite{Qin_2019,Kaminer_2011}. So, no matter different initial conditions, $k\rightarrow0$ at long time evolution. The stationary value of $\eta_s$ can be solved by $\dot{\eta_s}=0$ and get $\eta_s=\sqrt{3F_{\text{M}}/\left(2F_{\text{M}}-\Lambda\right)}$ with $F_{\text{M}}=\bar{R}\bar{\gamma}_{C}/\bar{\gamma}_{R}$ depending on the parameters of materials.

If the amplitude of the BIC solitons reaches the stationary value, the dynamics of the central position of the solitons can be solved by Eqs. (\ref{Dy2}) and (\ref{Dy3}). 
\begin{eqnarray}
k(t)=k(0)e^{2\Lambda\eta_s^2t/3},  z(t)=x_f-x_f e^{2\Lambda\eta_s^2t/3},
\end{eqnarray}
where $k(0)$ is the initial velocity of the soliton and $x_f$ is the final position. Notably, a negative value of $\Lambda$ corresponds to the exponential decay of the propagation. Consequently, during stable propagation within the condensates, the soliton trajectories undergo exponential deceleration and asymptotically approach a final stationary position.

As depicted in Figs. \ref{initialchange}(a3) and (b3), the final position of the solitons is highly sensitive to their initial conditions. Specifically, an increase in the initial amplitude $\eta(0)$ leads to an exponential decrease in the average final position $<x_f>$.  Conversely, a larger initial velocity $k(0)$ allows the soliton to propagate over a greater distance.

\begin{figure}
    \centering
    \includegraphics[width=1\linewidth]{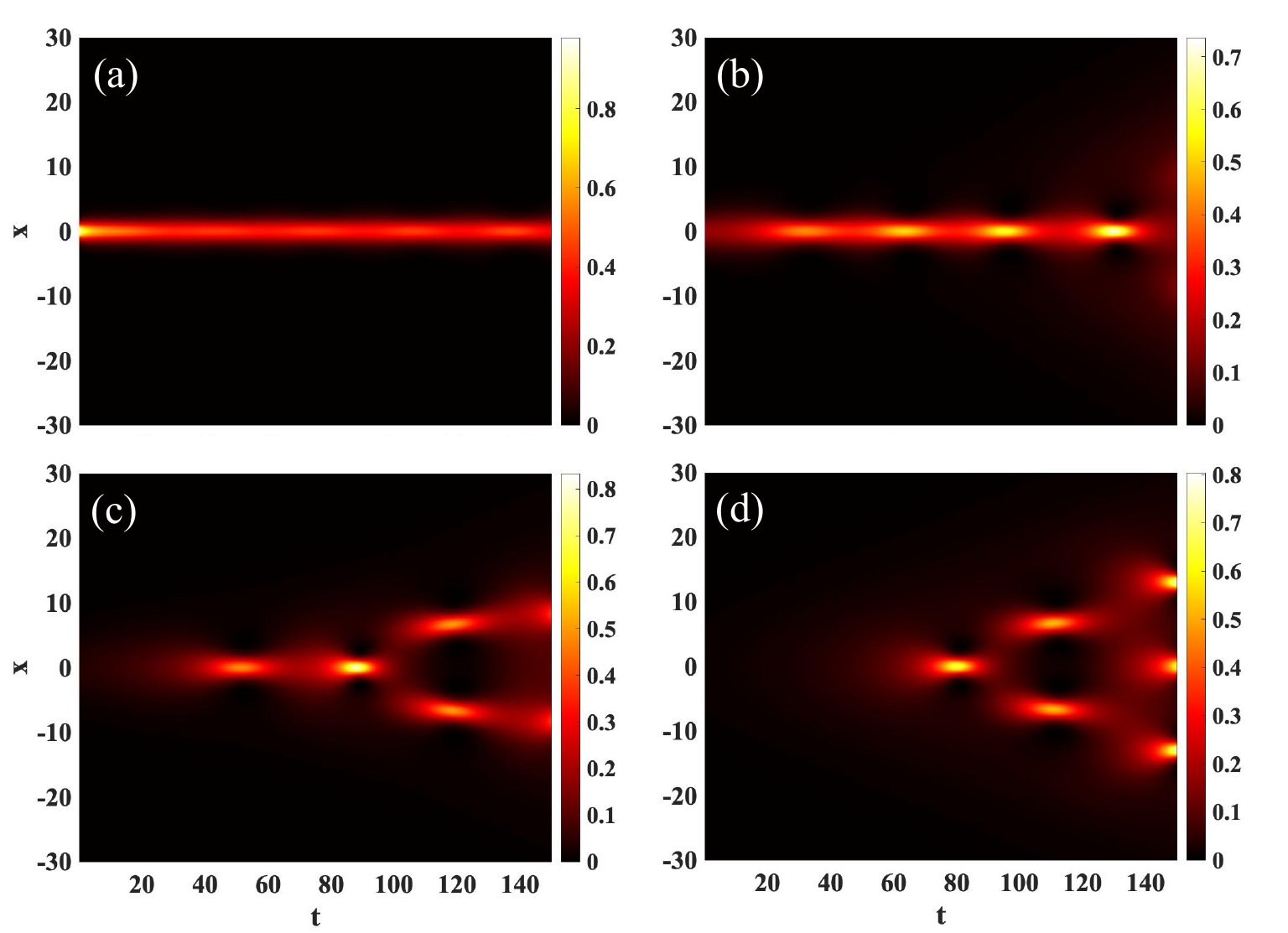}
    \caption{\textbf{Spatiotemporal density distribution  $|\psi(x,t)|^2$ of the BIC polaritons with different initial amplitudes.} We investigate the effect of the initial amplitudes with fixed parameters $k(0)=0$, and $\Lambda=-0.2$, and $\bar{\gamma}_R = 40$. The initial amplitudes of the bright solitons are $\eta(0) = 1$, 0.4, 0.2, 0.1 corresponding to subfigures (a), (b), (c), and (d). Other parameters are: $\bar{\gamma}_C=1$ and $\bar{R}=1.5$. \label{instability2} }
\end{figure}

The material-related parameters $\bar{\gamma}_R$ and $\Lambda$ also play a crucial role in the propagation of solitons, as illustrated in Figs. \ref{conditionchange}(a3) and (b3). Specifically, the BIC fitting parameter $\Lambda$ causes the solitons to decelerate exponentially as $|\Lambda|$ increases. When $\Lambda$ approaches zero, this deceleration vanishes, and the soliton exhibits normal dynamic behavior. In our variational ansatz, we assume $\bar{\gamma}_R\gg \bar{\gamma}_C$. If $\bar{\gamma}_R$ is not sufficiently large, the solitons fail to maintain stable trajectories. Conversely, $\bar{\gamma}_R$ is large enough to support stable solitons; the final position shows a linear relationship with $\bar{\gamma}_R$, as depicted in the inset of \ref{conditionchange}(b3).

\section{Dynamical Instability\label{secinstable}}

Exciton-polariton soliton stability is a critical factor governing their potential applications and robustness. The stability of these coherent structures is primarily maintained by a delicate balance between the inherent optical nonlinearity, stemming from exciton-exciton interactions, and wave diffraction, as theoretically established in driven-dissipative microcavities \cite{Woutersstable,Bobrovska2014}.  The interplay between gain and loss—governed by the complex interplay of exciton reservoirs and photon leakage—plays a decisive role in determining the stability phase diagram of both bright and dark solitons \cite{Ma_2026,Zhangstable}.  Here, we discuss how different parameters affect the dynamical stability of bright exciton-polariton solitons at the BIC by setting $k(0)=0$.

If we take $k(0)=0$, the central position of the soliton will remain at $z(t)=0$, and Eq. (\ref{Dy1}) can be analytically solved by 
\begin{equation}
\eta^2(t)=\frac{\eta_0^2}{\left[ 1+ (\frac{\Lambda}{3F_{\text{M}}} -\frac{2}{3})\eta_0^2\right]e^{-2F_{\text{M}} t}+\eta_0^2\left(\frac{2}{3}-\frac{\Lambda}{3F_{\text{M}}} \right)  }, \label{etats}
\end{equation}
where $\eta(0)=\eta_0$. If $\eta_0$ is very small, $\eta^2(t)$ exhibits an almost exponential expansion, leading to instability. Meanwhile, for a very large  $\eta_0$, $\eta(t)$ will have a stationary solution $\eta_s$. As illustrated in Figs. \ref{instability2} (a)-(d), along with the decrease of $\eta(0)$, the solitons become increasingly unstable and eventually split into multiple moving spots.

\begin{figure}
    \centering
    \includegraphics[width=1\linewidth]{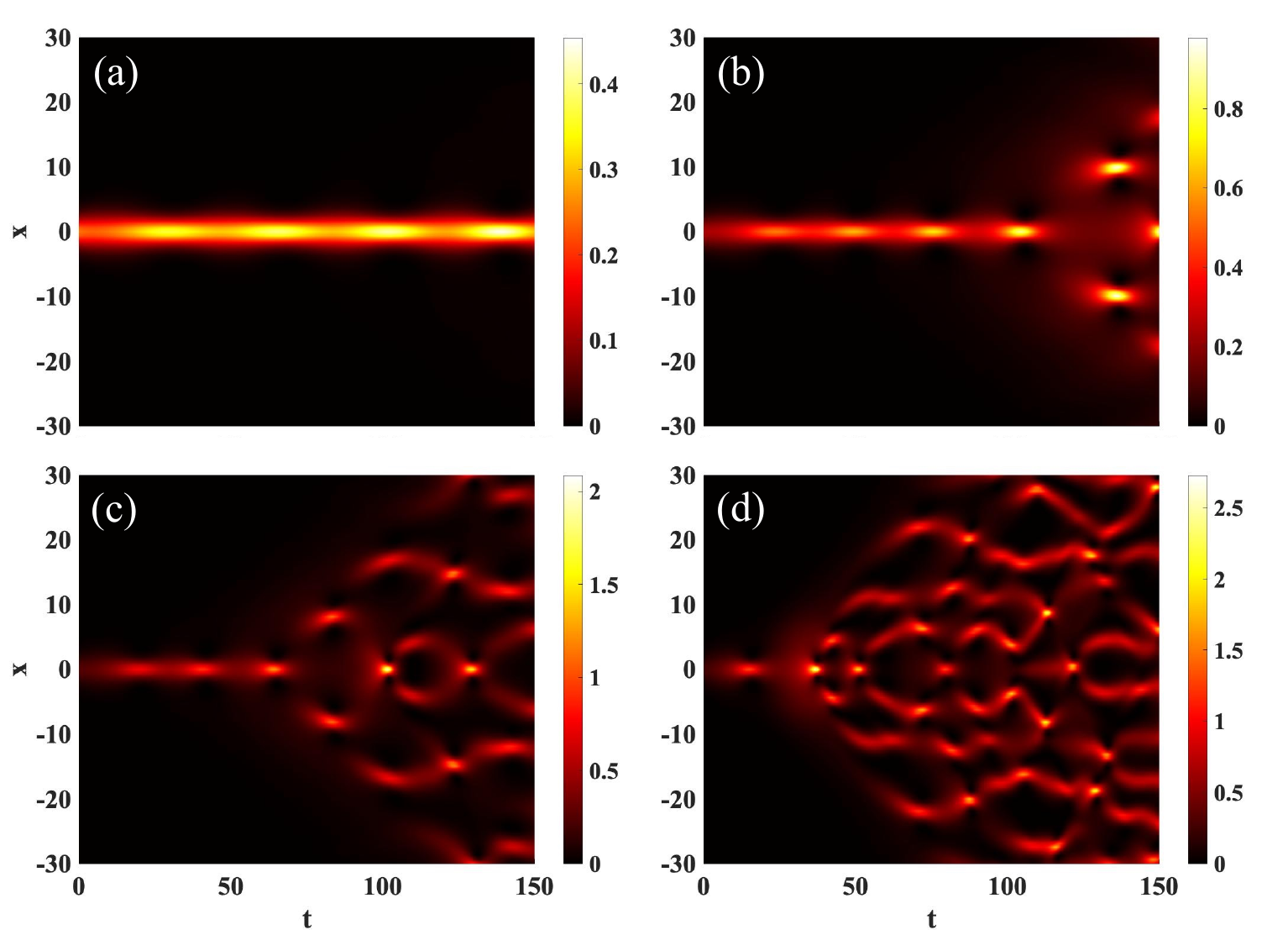}
    \caption{\textbf{Spatiotemporal density distribution  $|\psi(x,t)|^2$ of the BIC polaritons with different reservoir decay rates.} We investigate the effect of the reservoir decay rate with fixed parameters $\eta(0)=0.5$, $k(0)=0$, and $\Lambda=-0.2$. The reservoir decay rates are $\bar{\gamma}_R$ = 50, 30, 20, 10 corresponding to panels (a), (b), (c), and (d). Other parameters are: $\bar{\gamma}_C=1$ and $\bar{R}=1.5$.  \label{instability} }
\end{figure}

According to Eq. (\ref{etats}), the system-related parameter  $F_M$ will have a significant influence on the dynamical stability of the solitons. On one hand, the soliton amplitude evolves following an exponential function. On the other hand,  $F_{\text{M}}$ plays a critical role in controlling soliton stability. As shown in Fig. \ref{conditionchange}(b3), with fixed initial conditions of $\eta(0)=0.5$ and $k(0)=-0.4$, the system becomes unstable when $\bar{\gamma}_R<23$. Furthermore, when $k(0)=0$, the system is much more prone to instability as $\bar{\gamma}_R$ decreases; under these conditions, the solitons split into multiple stripes, as illustrated in Figs. \ref{instability}(a)-(d).

Remarkably, in our expression of $\eta(t)$, we formally take the limit of $k(t)=0$, which corresponds to the scenario where the soliton is initially at rest in the laboratory frame. However, contrary to the naive expectation that the initial velocity would become irrelevant in this limit, we find that the initial velocity $k(0)$ of the soliton can still profoundly influence the dynamical stability of the soliton during its subsequent evolution. Specifically, a higher initial velocity renders the soliton more robustly stable against perturbations, thereby allowing for significantly longer propagation distances before the onset of instability or dissipation. This counterintuitive result suggests that the memory of the initial kinematic state is retained in the nonlinear dynamics even when the apparent wave number vanishes. Furthermore, we emphasize that the BIC-fitted parameter $\Lambda$---which characterizes the effective non-Hermitian coupling strength in our model---must assume a negative value, i.e., $\Lambda < 0$, as a necessary condition for the existence of a physically meaningful steady state. Should $\Lambda$ become positive or vanish, the system would exhibit unbounded growth in the soliton amplitude, leading to a divergent behavior and consequently failing to reach any equilibrium or quasi-equilibrium configuration.

\section{Conclusion\label{secconclu}}

In conclusion, we have theoretically investigated the formation and nonequilibrium dynamics of bright exciton-polariton solitons in systems engineered with BIC. By employing a driven-dissipative Gross-Pitaevskii equation coupled with a rate equation for the excitonic reservoir, we demonstrate that BICs provide a robust and versatile platform for stabilizing polaritonic solitons against radiative decay, which conventionally constitutes a dominant loss channel in open photonic systems.

Utilizing a Lagrangian variational approach, we derive closed-form analytical expressions for the spatiotemporal trajectory and velocity of the solitons. Our results reveal that the propagation of BIC polariton solitons exhibits a distinct self-deceleration dynamics, eventually bringing the soliton to a complete halt at a finite final position. This terminal position is uniquely determined by the interplay between the initial momentum and the BIC fitting parameter $\Lambda$, which characterizes the effective non-Hermitian coupling arising from the engineered continuum. Specifically, a negative $\Lambda$ induces pronounced exponential deceleration, in stark contrast to the persistent or weakly damped dynamics typically observed in conventional polariton systems where radiative losses are present.

Furthermore, we have conducted a comprehensive analysis of the dynamical stability of these solitons against perturbations. We find that the system parameters, particularly the reservoir decay rate $\bar{\gamma}_R$ and the initial soliton amplitude $\eta(0)$, play crucial and nontrivial roles in maintaining solitonic integrity over extended propagation distances. While higher initial velocities are found to contribute favorably to stability—presumably due to enhanced kinetic energy overcoming dissipative perturbations—the system becomes increasingly prone to instability and catastrophic soliton splitting when the reservoir decay is insufficiently large or when the initial amplitude falls below a critical threshold. This identifies a delicate parameter window for experimental realization.

These findings highlight the exceptional potential of BIC-engineered structures for realizing stable, localized optical excitations with controlled and tunable mobility. Our work offers valuable insights and design principles for applications in low-threshold polariton lasers, where suppressed radiative losses may enable significantly reduced pumping requirements, as well as in topological photonics, where robust localized states are essential for protected information transport.

\section{Acknowledgement}
The work is supported by the National Natural Science Foundation of China (Grant No. 12404362) and the Fundamental Research Funds for the Central Universities (Grant No. JUSRP123027).
\bibliography{mybib}

\end{document}